\documentstyle{lamuphys}
\input{psfig}
\makeatletter
\let\chapter\hid@chapter
\makeatother
\begin{document}
\pagenumbering{arabic}
\title{The Deep X-ray Radio Blazar Survey (DXRBS)}

\author{Paolo\,Padovani\inst{1,2,3}, Eric\,Perlman\inst{1}, 
Paolo\,Giommi\inst{4}, Rita\,Sambruna\inst{5}, Laurence R.\,Jones\inst{6},
Anastasios\,Tzioumis\inst{7} and John\,Reynolds\inst{7}}

\institute{Space Telescope Science Institute, 3700 San Martin Drive, 
Baltimore MD. 21218, USA
\and
Affiliated to the Astrophysics Division, Space Science Department, European
Space Agency
\and
On leave from Dipartimento di Fisica, II Universit\`a di Roma ``Tor Vergata'',
Italy
\and
SAX Science Data Center, ASI, Viale Regina Margherita 202, 
I-00198, Italy
\and
Pennsylvania State University, Department of Astronomy, 525 Davey Lab,
University Park, PA 16803
\and
School of Physics \& Astronomy, Univ. of Birmingham,
Birmingham B15 2TT, UK
\and
Australia Telescope National Facility, CSIRO, PO Box 76, Epping NSW
2121, Australia}
\authorrunning{Padovani et al.}
\maketitle
\
\begin{abstract}

We have undertaken a survey for blazars by correlating the ROSAT WGACAT
database with publicly available radio catalogs, restricting our
candidate list to serendipitous flat-spectrum sources ($\alpha_{\rm r} \le
0.7$, $f_{\nu} \propto \nu^{-\alpha}$). We discuss here our survey methods,
identification procedure and first results. Our survey is found to be
$\sim 95\%$ efficient at finding blazars, a figure which is
comparable to or greater than that achieved by other radio and X-ray survey
techniques. DXRBS provides a much more uniform coverage of the parameter space
occupied by blazars than any previous survey. Particularly important is the
identification of a large population of flat-spectrum radio quasars with
ratios of X-ray to radio luminosity $\ga 10^{-6}$ ($\alpha_{\rm rx} \la 0.78$)
and of many low-luminosity flat-spectrum radio quasars. Moreover, DXRBS fills
in the region of parameter space between X-ray selected and radio-selected
samples of BL Lacs.

\end{abstract}
\section{Introduction}
Blazars are the most extreme variety of Active Galactic Nuclei (AGN) known.
Their signal properties include irregular, rapid variability; high optical
polarization; core-dominant radio morphology; apparent superluminal motion;
flat ($\alpha_{\rm r} \la 0.5$) radio spectra; and a broad continuum extending
from the radio through the gamma-rays (e.g., Urry \& Padovani 1995). The
broadband emission from blazars is dominated by non-thermal processes (most
likely synchrotron and inverse-Compton radiation). Blazar properties are
consistent with relativistic beaming, that is bulk relativistic motion of the
emitting plasma at small angles to the line of sight (as originally proposed
by Blandford \& Rees in 1978), which gives rise to strong amplification and
collimation in the observer's frame. It then follows that an object's
appearance depends strongly on orientation. Hence the need for ``Unified
Schemes'', which look at intrinsic, isotropic properties, to unify
fundamentally identical (but apparently different) classes of AGN.

The blazar class includes flat-spectrum radio quasars (FSRQ) and BL Lacertae
objects. The main difference between the two classes lies in their emission
lines, which are strong and quasar-like for FSRQ and weak or in some cases
outright absent in BL Lacs. 
As a consequence of their peculiar orientation with respect to our line of
sight, blazars represent a very rare class of objects, making up considerably
less than 5\% of all AGN (Padovani 1997).


\section{Why a Blazar Survey?}
A blazar survey is needed essentially for two reasons: number statistics and
limiting fluxes. All existing blazar samples, in fact, are relatively small
and at relatively high fluxes. The largest radio-selected BL Lac sample, the 1
Jy sample (Stickel et al. 1991), includes 37 objects with $f_{\rm 5 GHz}
> 1$ Jy. In the X-rays, there are a few available samples, made up of $40 -
70$ objects (the EMSS sample [Maccacaro et al. 1994], the Slew sample [Perlman
et al. 1996], plus various ROSAT-based samples [Laurent-Muehleisen et
al. 1997; Bade et al., in preparation]) but the deepest sample is still the
EMSS one, which reaches $f_{\rm x} \sim 2 \times 10^{-13}$ erg cm$^{-2}$ 
s$^{-1}$. (See however Wolter et al. 1997 for an ongoing deeper survey.)

Moving to FSRQ, until very recently the only sizeable sample with complete
redshift information was the one extracted from the 2 Jy sample (Wall \&
Peacock 1985), which includes 52 sources.
(The FSRQ in the 1 Jy and S4 samples reach lower fluxes [1 Jy and 0.5 Jy
respectively] and are more numerous but the identification of the two samples
is still not complete.) Drinkwater et al. (1997) have recently published the
PKS 0.5 Jy sample, which includes 323 flat-spectrum radio sources with $f_{\rm
2.7 GHz} > 0.5$ Jy, 86\% of which have a measured redshift. Finally, Shaver,
Hook, and collaborators have just put together a sample of more than 400 
flat-spectrum objects down to $f_{\rm2.7 GHz} = 0.25$ Jy (Hook, these 
proceedings).

There is then clearly a need for a deeper, larger blazar survey, to address
many open questions of blazar research, namely:

\begin{itemize}

\item BL Lac evolution and luminosity functions; unified schemes

\item Flat-spectrum radio quasar evolution and luminosity functions; unified 
schemes

\item The relationship between BL Lacs selected in the radio and X-ray band

\item The relationship between flat-spectrum radio quasars and BL Lacs

\item What is a BL Lac?

\item The relationship between physical parameters (e.g., X-ray and
radio spectral indices, equivalent width, line luminosity, continuum
luminosity, etc.)

\end{itemize}

\section{The DXRBS}

The basic idea behind our Deep X-ray Radio Blazar Survey (DXRBS) is quite
simple: blazars are relatively strong X-ray and radio emitters so selecting
X-ray and radio sources with flat radio spectrum (one of their defining
properties) should be a very efficient
way to find these rare sources. We adopt a spectral index cut $\alpha_{\rm r}
\le 0.7$. This will: 1. select (by definition!) flat-spectrum radio quasars;
2. select basically 100\% of BL Lacs; 3. exclude the large majority of radio
galaxies. 

We have then cross-correlated WGACAT (White, Giommi \& Angelini 1995), the
publicly available database of ROSAT PSPC sources (restricting ourselves to
sources having quality flag $\ge 5$ to avoid problematic detections) with a
number of publicly available radio catalogs. North of the celestial equator,
we used the 20 cm and 6 cm Green Bank survey catalogs NORTH20CM and GB6 (White
\& Becker 1992; Gregory et al. 1996), while south of the equator, we used the
Parkes-MIT-NRAO catalog PMN (Griffith \& Wright 1993). All sources with radio
spectral index $\alpha_{\rm r} \leq 0.7$ at a few GHz were selected as blazar
candidates. Note that WGACAT reaches $f_{\rm x} \sim 10^{-14}$ erg cm$^{-2}$
s$^{-1}$ (although its flux limit varies widely on the sky), while the flux
limits of the radio catalogs are the following: $f_{\rm 5 GHz} \sim 20$ mJy
(GB6), $f_{\rm 1.4 GHz} \sim 100$ mJy (NORTH20CM), and $f_{\rm 5 GHz} \sim 40$
mJy (PMN). (See Fig. 1 and section 4.)

For objects north of the celestial equator, $6-20$ cm radio spectral indices
were obtained directly from the cross-correlation of the GB6 and NORTH20CM
catalogs. For sources at southern declinations, the lack of a comparably deep
radio survey at a second frequency required a different strategy\footnote{The
NVSS survey (Condon et al. 1997) was not available when we started this
project. Moreover, it is still not 100\% completed and covers the sky north of
$-40^\circ$.}. We then conducted a snapshot survey with the Australia Telescope
Compact Array (ATCA) at 3.6 and 6 cm, to get also radio spectral indices
unaffected by variability (which will be a problem for our northern sample).

As a result of the correlations between the X-ray and the radio catalogs we
have obtained a list of about 200 blazar candidates (with $|b| > 10^{\circ}$),
to which we add 88 previously known, serendipitous (i.e., not ROSAT targets)
blazars (77 FSRQ and 11 BL Lacs), for a total of about 300 sources. 

We note that, as the original catalogs included tens of thousands of objects,
our search strategy has narrowed down the number of candidates by more than
two orders of magnitude. This kind of approach is extremely important for
surveys, like ours, that look for rare objects in large catalogs and will
be vital with the advent of even larger and deeper catalogs, foreseen in the
near future. 

\subsection{The Identifications}

Accurate positions to pinpoint the optical counterparts were
obtained from either the NVSS (Condon et al. 1997) or our ATCA survey.
Magnitudes for all X-ray/radio sources with counterparts on the POSS and UKST
plates which comprise the Digitized Sky Survey were obtained from the
Cambridge APM and Edinburgh COSMOS projects (Irwin et al. 1994; Drinkwater et
al. 1995). Most X-ray/radio sources without counterparts on the survey plates
were imaged at either the KPNO 0.9m or the CTIO 0.9m telescopes. This allowed
identification of optical counterparts to $R = 23$. The magnitude distribution
of the blazar candidates peaks around 18.  Spectroscopic observations were
conducted at the KPNO 2.1 m, MMT, Lick 3 m, ESO 2.2 m and 3.6 m, and CTIO 1.5
m telescopes.

\begin{table}[htb]
\caption[ ]{DXRBS Identifications}
\begin{flushleft}
\renewcommand{\arraystretch}{1.2}
\begin{tabular}{lrrr}
\hline\noalign{\smallskip}
Class & ~~~Newly identified& ~~~Previously known& ~~~Total \\
\noalign{\smallskip}
\hline
\noalign{\smallskip}
Radio Quasars & 86 & 77 & 163 \\
BL Lacs & 26 & 11 & 37 \\
Radio Galaxies & 4 & 10 & 14 \\
\noalign{\smallskip}
\hline
Total & 116 & 98 & 214 \\
\noalign{\smallskip}
\hline

\end{tabular}
\renewcommand{\arraystretch}{1}
\end{flushleft}
\end{table}

The breakdown of the identifications at the time of writing (February 1998) is
given in Table 1. So far we have identified $\sim 50\%$ of our candidates and
97\% of them are blazars. 90\% of the previously known objects are also
blazars\footnote{Note that it can be difficult to distinguish between a BL Lac
and a radio galaxy for border-line sources. Therefore, while we are confident
that the classification criteria have been applied consistently for our newly
discovered blazars, this might not be the case for the previously known
sources.}.  Our method is then indeed very efficient (93.5\%) at identifying
blazars.

\begin{figure}
\psfig{file=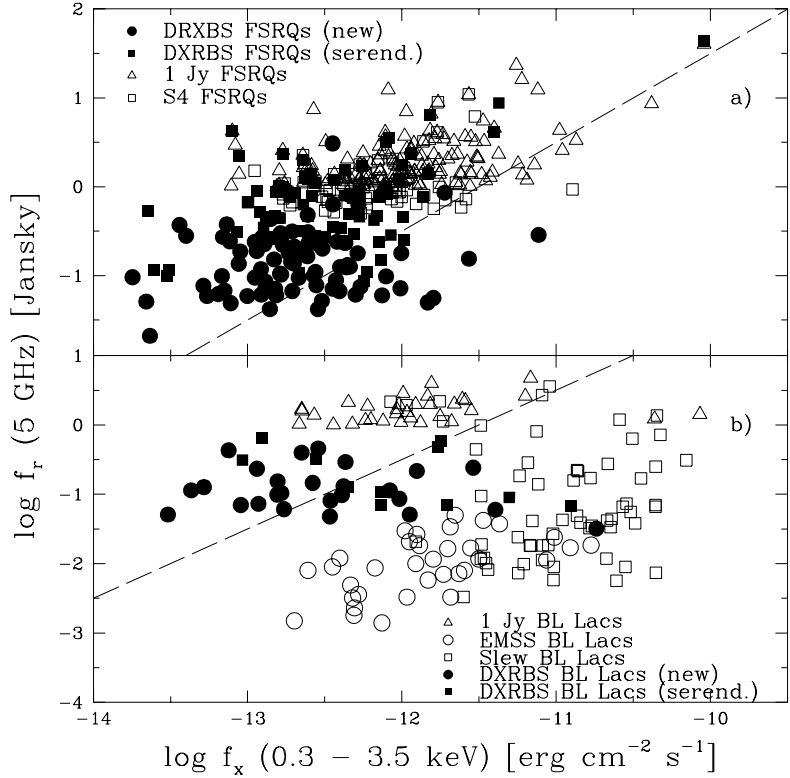,width=11.0truecm}
\caption[ ]{The radio/X-ray flux plane for FSRQ (a); top) and BL Lac (b);
bottom) samples. Note how DXRBS is sampling previously unexplored regions of
parameter space. The dashed lines correspond to $\log f_{\rm x}/f_{\rm r} =
-11.5$.}
\end{figure}

\section{First Results}
The 163 DXRBS FSRQ we have so far indentified span the redshift range of 0.1
to 3.8, with a mean value of 1.25. Figure 1a shows the greatly improved
coverage of the radio/X-ray flux plane for FSRQ provided by DXRBS, as compared
to previously available samples (note that both the 1 Jy and S4 samples are
not completely identified and X-ray data are available only for $\sim 60\%$ of
their FSRQ). DXRBS FSRQ go about an order of magnitude deeper in radio flux
and a factor of $\sim 4$ deeper in X-ray flux. Moreover, while
very few previously known FSRQ had relatively large X-ray-to-radio flux
ratios ($f_{\rm x}/f_{\rm r} > 10^{-11.5}$ in the units of Fig. 1: see the 
dashed line), many DXRBS FSRQ are quite ``X-ray bright.''

\begin{figure}
\psfig{file=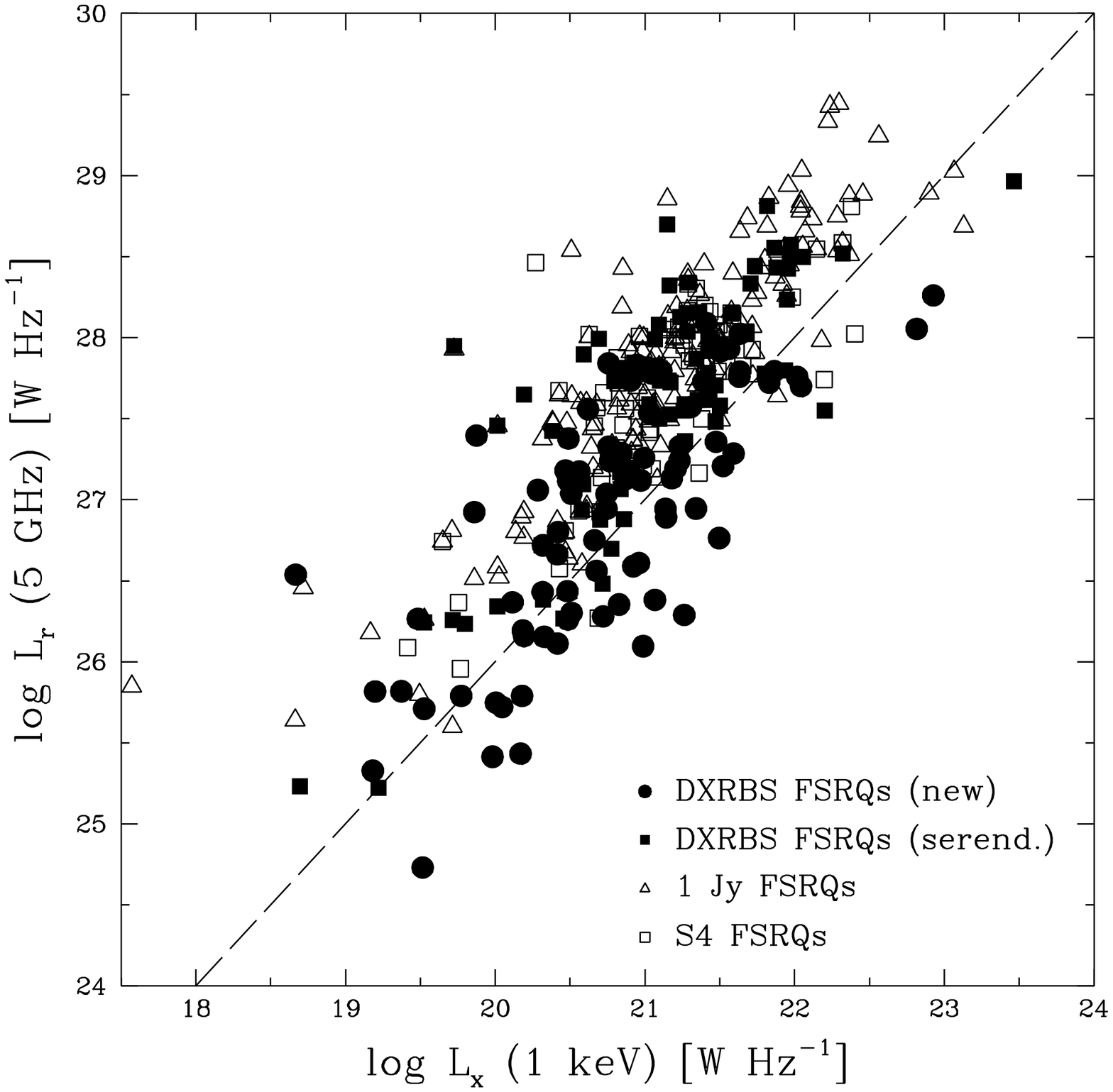,height=9.1truecm,width=10.0truecm}
\caption[ ]{The radio and X-ray luminosities of FSRQ.} 
\end{figure}

This is better seen in Fig. 2, which plots radio versus X-ray luminosity for
FSRQ. About 25\% of DXRBS FSRQ have $L_{\rm x}/L_{\rm r} \ga 10^{-6}$ (or,
alternatively, $\alpha_{\rm rx} \la 0.78$). Only nine 1 Jy/S4 FSRQ had such
luminosity ratios. Based on the overall spectral energy distribution of these
sources and extrapolating from the situation in BL Lacs, these ``X-ray
bright'' FSRQ probably have the peak of their synchrotron emission at UV/X-ray
energies, unlike the other, more common, FSRQ, which peak in the IR/optical
band. Due to their lower radio fluxes, DXRBS FSRQ are also reaching relatively
low radio luminosities, approaching what should be the lower end of the FSRQ
luminosity function according to unified schemes ($L_{\rm r} \approx 10^{24.5}$
W Hz$^{-1}$: Urry \& Padovani 1995). In particular, more than 20\% of them have
$L_{\rm r} < 10^{26.5}$ W Hz$^{-1}$, as compared to only 3\% for the 1 Jy and
S4 samples.

As regards BL Lacs, DXRBS is again exploring uncharted territory, as it is
finding BL Lacs which cover a previously unexplored region of the radio/X-ray
plane, and go deeper (by almost an order of magnitude) in X-ray flux than
currently available samples (see Fig. 1b). Note also how the DXRBS BL Lacs are
intermediate in their X-ray-to-radio flux ratios as compared to the
``classical'' radio and X-ray selected samples. A more complete description of
the DXRBS first results is given by Perlman et al. (1998).

%

%
%

\end{document}